\documentclass[aps,prb,twocolumn,amsmath,amssymb,superscriptaddress,floatfix]{revtex4}
\usepackage{graphicx}
\usepackage{bm}
\usepackage[usenames]{color}
\bibstyle{apsrev.bib}

\newcommand{\be}{\begin{equation}}
\newcommand{\ee}{\end{equation}}
\newcommand{\beqn}{\begin{eqnarray}}
\newcommand{\eeqn}{\end{eqnarray}}

\begin{document}

\title{Entanglement entropy dynamics of disordered quantum spin chains}

\author{Ferenc Igl\'oi}
\email{igloi.ferenc@wigner.mta.hu}
\affiliation{Wigner Research Centre, Institute for Solid State Physics and Optics,
H-1525 Budapest, P.O.Box 49, Hungary}
\affiliation{Institute of Theoretical Physics,
Szeged University, H-6720 Szeged, Hungary}
\author{Zsolt Szatm\'ari}
\email{zsolt.szatmari@me.com}
\affiliation{Institute of Theoretical Physics,
Szeged University, H-6720 Szeged, Hungary}
\author{Yu-Cheng Lin}
\email[Corresponding author: ]{yc.lin@nccu.edu.tw}
\affiliation{Graduate Institute of Applied Physics, 
National Chengchi University, Taipei, Taiwan}
\date{\today}


\begin{abstract}
By means of free fermionic techniques we study the time evolution of the
entanglement entropy, ${\cal S}(t)$, of a block of spins in the random
transverse-field Ising chain after a sudden change of the parameters of the
Hamiltonian. We consider global quenches, when the parameters are modified
uniformly in space, as well as local quenches, when two disconnected blocks are
suddenly joined together.  For a non-critical final state, the dynamical
entanglement entropy is found to approach a finite limiting value for both
types of quenches. If the quench is performed to the critical state,
the entropy grows for an infinite block as ${\cal S}(t) \sim \ln \ln t$.  This
type of ultraslow increase is explained through the strong disorder
renormalization group method.
\end{abstract}

\maketitle
\section{Introduction}
\label{sec:intr}
Recently, we have witnessed increasing interest in studying the entanglement
properties of quantum many body systems\cite{amico,entanglement_review,area} in
different disciplines: quantum information, condensed matter physics and
quantum field theory. Among the various measures for quantifying entanglement,
the von Neumann entropy and its generalizations to R\'enyi entropies, as well as
the entanglement spectrum\cite{Calabrese_Lefevre} have been widely used to
obtain useful information about the topological and universal properties of an
extended quantum system, in particular at a quantum critical point. For
homogeneous, i.e. non-random, systems, many basic results are known in one
dimension from conformal field theory,\cite{holzhey,Calabrese_Cardy04}
which have been confirmed by exact and numerical calculations on specific
models.\cite{vidal,peschel03,jin_korepin,peschel04,IJ07}  For a quantum spin
chain one generally considers the entanglement entropy, ${\cal S}_{\ell}$,
between a block of $\ell$ contiguous spins and its complement. For periodic
chains, where the block has two boundary points connected with the remainder of the system,
the entanglement entropy at the critical point for $\ell \gg 1$ scales as
${\cal S}_{\ell}=\frac{c}{3}\ln \ell$, where $c$ is the central charge of the
conformal field theory. Away from the critical point, the entropy saturates to a
value $ {\cal S}_{\ell}=\frac{c}{3}\ln \xi$, where $\xi \ll \ell$ (and $\xi
\gg 1$) is the correlation length of the system.  Recently universal
finite-size corrections to the R\'enyi entropy \cite{LSCA,CCEN,CC10,ABS} as well as the
entropy of non single-connected blocks have also been
studied.\cite{ATC,igloi_peschel10}

If the couplings in the chain are inhomogeneous, such as there is an internal
defect\cite{peschel03,PeschelZhao,Levine08,ISzL09,Eisler_Peschel10,CMV11,peschel12} or the
interactions are quasi-periodic or aperiodic,\cite{IJZ07} then the prefactor of
the critical entanglement entropy, the so called effective central charge,
$c_{\text{eff}}$, is generally different from that in the homogeneous system.
For chains with random couplings, $c_{\text{eff}}$ has been calculated
analytically\cite{refael,Santachiara,Bonesteel,s=1} by the  strong
disorder renormalization group (SDRG) method,\cite{im} and numerically by free-fermionic
techniques\cite{Laflo05,IgloiLin08} and by the density-matrix renormalization
group (DMRG) method.\cite{dyn06} Also the entanglement spectrum of random XX chains
has been studied, both by the SDRG method and numerically.\cite{FCM11} We note that the entanglement entropy can be
studied even in higher dimensional random quantum systems by  numerical
implementation of the SDRG method,\cite{lin07,yu07,sdrg_entr} provided the
critical properties of the systems are controlled by infinite-disorder fixed
points,\cite{danielreview,im} as in one dimension.

The nonequilibrium quench dynamics of quantum systems has become a very active
field of research, both experimentally and theoretically.\cite{quench_rev}
Dynamical aspects of the entanglement entropy are  of interest for their close
relationship to the speed of information propagation through an interacting
quantum system. In these investigations  one changes (some) parameters of the
Hamiltonian suddenly, and asks how the entanglement evolves in time.\cite{CC05}
One generally distinguishes two types of quenches: global and local quenches.
For a global quench the parameters are changed everywhere in space. In this case
the entanglement entropy has a linear increase in time $t$, irrespective of the
initial and the final state of the system. This type of dynamics has been
explained in terms of quasiparticles (elementary excitations).\cite{CC05} In
the other type of quench, known as a local quench, the parameters of the
Hamiltonian are changed only locally; for example, a block, which is
disconnected from the rest of the system for $t<0$, is instantaneously
connected at time $t=0$.  For the local quench the entanglement entropy at the
critical point is found to display a universal logarithmic
increase,\cite{EP07} ${\cal S}_{\ell}=\frac{2c}{3}\ln t$, $t \ll \ell$; this
relation has been later derived through conformal
invariance.\cite{CC07,stephan_dubail}

Concerning dynamical entropy in inhomogeneous systems there have been only a
few studies in specific situations. \cite{dyn06,ISzL09,burrell_osborne,lieb_robinson,further_dyn} For local
quench the effect of defects has been studied. The defects can be, for
example, in the form of couplings between a block, where the entanglement
entropy is studied, and the rest of system; for this case, the prefactor of the
logarithmic $t$-dependence of the entanglement entropy is found to be the same
as measured in the static case. \cite{ISzL09} On the other hand, quenched
disorder changes the entanglement dynamics in a more drastic way.  Previously,
the entanglement entropy dynamics of the disordered Heisenberg chain following
a global quench was numerically studied,\cite{dyn06} where a slow increase of
the entropy with time was observed; the numerical data in the time regime
($t\lesssim 500$) obtained by time dependent DMRG suggested that the entropy
grows logarithmically with time. The slow propagation of signals in
disordered system has been later supported by theoretical
work,\cite{burrell_osborne} in which, by means of the generalized  Lieb-Robinson
bound,\cite{lieb_robinson} a bound for time evolution of the entanglement
entropy is derived in the form: ${\cal S}_{\ell}(t) \le c_1+c_2 \log(\ell|t|)$,
with $c_1$ and $c_2$ being constants.    

In this paper we revisit the problem of the entanglement entropy dynamics in
disordered quantum spin chains. The model we consider is the random
transverse-field Ising chain. Our study extends previous investigations on
entanglement dynamics in disordered systems in several respects: (i) we study
the entanglement dynamics both at the critical point and in the off-critical
phases, using numerically exact free-fermionic techniques; (ii) we consider
both global and local quenches; (iii) we study the time evolution for a very long
period of time and obtain the long-time asymptotics for finite systems; (iv)
furthermore, we explain the numerical findings based on SDRG.

The structure of the rest of the paper is as follows. In section
\ref{sec:model} we introduce the model, describe its basic equilibrium
properties and outline the method of calculation. Results of the entanglement
entropy dynamics after global and  local quenches are presented in sections
\ref{sec:global} and \ref{sec:local}, respectively.  The results are discussed
in section \ref{sec:disc}.

\section{The model}
\label{sec:model}

The model we consider is the quantum Ising chain of length $L$ defined by the Hamiltonian:
\be
{\cal H}=-\sum_{i=1}^{L} {J}_i \sigma_i^x \sigma_{i+1}^x -\sum_{i=1}^{L} {h}_i \sigma_i^z\;,
\label{hamilton}
\ee
in terms of the Pauli matrices $\sigma_i^{x,z}$ at site $i$.  In this paper we
will take periodic boundary conditions so that $\sigma_{L+1}=\sigma_1$.  The
homogeneous model with the couplings $J_i=1$ and the transverse fields
$h_i=\tilde{h}$ is in the disordered (ordered) phase for $\tilde{h}>1$
($\tilde{h}<1$), and the quantum critical point is located at
$\tilde{h}=1$.\cite{pfeuty} The critical point of the model is described by a
conformal field theory with a central charge $c=1/2$. In the random
model with quenched disorder, the ${J}_i$ and the ${h}_i$ are position
dependent, and are independent random numbers taken from uniform distributions
in the intervals $[0,1]$ and $[0,1]h$, respectively. The random model is
in the disordered (ordered) phase for $h>1$ ($h<1$) and the random quantum critical
point is at $h=1$. The equilibrium critical properties of the random chain has
been studied by the SDRG method,\cite{fisher} and the random quantum critical
point is found to be controlled by an infinite-disorder fixed point, at which
the scaling is extremely anisotropic, so that the typical length, $\xi$, and
the typical time, $\tau$, is related as 
\be
\ln \tau \sim \xi^{\psi}\;, 
\label{psi}
\ee
with an exponent $\psi=1/2$.

In this work we study the entanglement entropy of a block of contiguous spins
sitting on sites $1,2,\cdots, \ell$ in the chain; the entanglement entropy is
defined as  ${\cal S}_{\ell}={\rm Tr}_{\ell}[\rho_{\ell} \ln\rho_{\ell}]$ in
terms of the reduced density matrix: ${\bf \rho}_{\ell}={\rm
Tr}_{i>\ell}|0\rangle\langle 0 |$, where $|0\rangle$ denotes the ground state
of the complete system with $L$ sites. In the calculation we make use of the
fact that the Hamiltonian in Eq.(\ref{hamilton}) can be expressed in terms of
free fermions,\cite{lsm,pfeuty} and the density matrix of the corresponding
free fermionic system is then obtained from its correlation
matrix.\cite{peschel03,vidal,IJ07} This calculation for a system in equilibrium
is straightforward. In the nonequilibrium case with quench dynamics one has two
Hamiltonians, say ${\cal H}_0$ for $t<0$ and ${\cal H}$ for $t>0$, both in the
form of Eq.(\ref{hamilton}) but with different parameters. The time-evolution
of the density matrix is governed by ${\cal H}$, as ${\bf
\rho}(t)=\exp(-\imath{\cal H}t) {\bf \rho} \exp(\imath{\cal H} t)$, but its
matrix elements are calculated through the eigenstates of the initial
Hamiltonian ${\cal H}_0$. Details of the calculation of the dynamical entropy
can be found in Ref.~\onlinecite{ISzL09}.

In the following, we first present results for global quenches and then for local
quenches. In each case, we will first briefly discuss the homogeneous model to compare
with our main results for the random chain, which will be given subsequently.

\section{Global quench}
\label{sec:global}
In a global quench the parameters of the Hamiltonian are modified everywhere in space. Concerning the
Hamiltonian in Eq.~(\ref{hamilton}), we modify the transverse fields, but leave the couplings unaltered
in the quench procedure.

\subsection{Homogeneous chain}
\label{sec:global_hom}

\begin{figure}[h!]
\begin{center}
\includegraphics[width=8cm, clip]{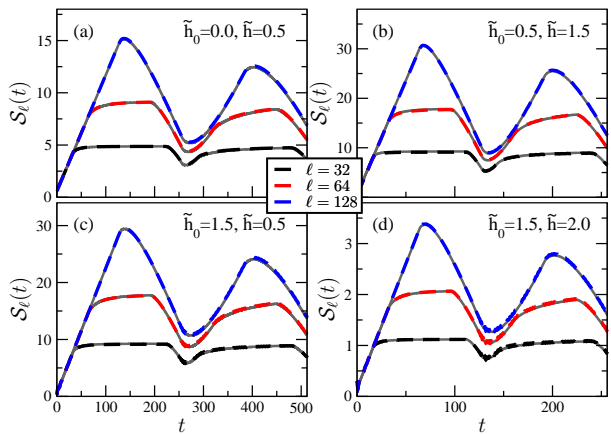}
\caption{(Color online) 
Entanglement entropy of a block of $\ell$ contiguous spins in the homogeneous
chain of length $L=256$ after a quench from a state with $\tilde{h}_0$ to a state with
$\tilde{h}$. Results of the semiclassical calculation are given by the gray curves.}
\label{fig1}
\end{center}

\end{figure}

In the homogeneous chain the transverse fields are changed from ${\tilde h}_0$ to ${\tilde h}$ and
we measure ${\cal S}_{\ell}(t)$ in a chain of total length $L=256$ for various sizes
of the block $\ell$. Different combinations of ${\tilde h}_0$ and ${\tilde h}$ are considered,
including quenches from an ordered state to another ordered state [Fig.~\ref{fig1} (a)], from
a disordered state to another disordered state [Fig.~\ref{fig1} (d)] and quenches 
through the critical point [Fig.~\ref{fig1} (b) and \ref{fig1} (c)].  

In each combination of ${\tilde h}_0$ and ${\tilde h}$ the dynamical entropy
has a similar behavior.  After a linearly increasing period, ${\cal
S}=\alpha({\tilde h}_0,{\tilde h}) t$, the entropy saturates to a value ${\cal
S}=\beta({\tilde h}_0,{\tilde h}) \ell$, and decreases subsequently. This time
dependence of ${\cal S}(t)$ repeats quasi-periodically, which is different
from the case in the thermodynamic limit $L\to\infty$, where the entropy
remains constant after it saturates. \cite{CC05,Fagotti08}

We recall that the exact values of $\alpha({\tilde h}_0,{\tilde h})$ and
$\beta({\tilde h}_0,{\tilde h})$ for $L\to\infty$ and for a large $\ell$ have
been calculated; \cite{Fagotti08} for the special case, when the quench is
performed to the critical state with ${\tilde h}=1$, the coefficient
$\alpha({\tilde h}_0,{\tilde h}=1)$ can be calculated in a closed
form.\cite{eip09} Following a semiclassical approach in terms of ballistically
moving quasiparticles formulated in Ref.~\onlinecite{semi_class}, we are able
to calculate the dynamical entropy for the finite chain. The quasiparticles are
Fourier transforms of kink states, created as a pair of entangled
free fermions with quasimomenta $\pm p$. \cite{CC05,semi_class} With energy
$\epsilon_p$, a ballistically moving quasiparticle has the semiclassical
velocity  $v_p=\frac{\partial \epsilon_p}{\partial p}$. These quasiparticles
are created homogeneously in space at $t=0$ with an occupation probability
$f_p$. If a pair of entangled particles arrive simultaneously in the block and
outside the block, they will contribute $s_p=-(1-f_p)\ln(1-f_p)-f_p\ln f_p$
to the entanglement entropy.  Summing up the contributions from all
quasiparticles (i.e. over positions and quasimomenta) we obtain the dynamical
entropy as shown in Fig.~\ref{fig1} by the grey curves.  As can be seen, the results
obtained from this semiclassical approach fit our numerically exact data
perfectly, even for a large range of $t$.

\subsection{Random chains}
\label{sec:global_rand}
Now we turn to the disordered chain with random couplings and random fields.
In our quench procedure, the set of couplings $\{J_i\}$ for a given sample
remains unaltered, whereas the width of the transverse-field distribution is
changed from $h_0$ ($t<0$) to $h$ ($t\ge0$). We consider quenches in the off-critical
phases as well as quenches to the critical point. In most of our calculations
the widths $h_0$ and $h$ are different; we keep the local transverse field
on each site correlated before and after the quench, but change
the relative magnitude by $h_i(t<0)/h_i(t\ge0)=h_0/h, \forall i$.
We also consider the case where the quench is performed at the critical
point, i.e. $h_0=h=1$; for this case we use two independent sets of
random variables for the transverse fields before and after the quench.  

We have calculated the disorder-averaged entanglement entropy between two
blocks of length $\ell=L/2$ in a chain for different system sizes up to
$L=256$. We have used at least 10,000 disordered realizations to obtain the
disorder average.

The time-dependence of the average entanglement entropy for quenches to a
non-critical state and for quenches to the critical point is qualitatively
different. Below we present results for these two cases separately.

\subsubsection{Quench to non-critical states}
\label{sec:global_rand_out}
\begin{figure}
\begin{center}
\includegraphics[width=8cm, clip]{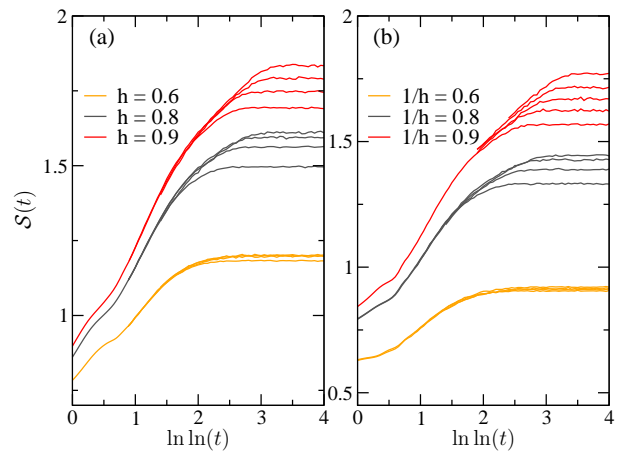}
\caption{(Color online) Disorder-averaged entanglement entropy for  several global quenches
performed outside the critical point of random chains, plotted against double
logarithmic time $\ln \ln t$. (a): quench from a fully ordered initial state
with $h_0\to 0$ to an ordered state with $h=0.6,~0.8$ and $0.9$; (b): quench from
a fully disordered initial state with $1/h_0\to 0$ to a disordered state with
$1/h=0.6,~0.8$ and $0.9$. System sizes ranging from $L=32$ to $L=192$ are
considered; the bigger the system, the higher the value of the entropy at large
$t$.}
\label{fig2}
\end{center}

\end{figure}

We performed quenches from a fully ordered initial state $h_0\to0$
to ordered states with $0<h<1$ and quenches from a fully
disordered initial state $h_0\to\infty$ to disordered states
with finite $h>1$. 
Results for the disorder-averaged entanglement entropy 
as a function of time for quenches in the ordered phase 
with parameters $h=0.6$, $h=0.8$ and $h=0.9$
are shown in Fig.\ref{fig2}(a). Results for
quenches in the disordered phase with parameters
$1/h=0.6$, $1/h=0.8$ and $1/h=0.9$ are shown in Fig.\ref{fig2}(b).  
In both cases the time variation is extremely slow, therefore we have 
used double-logarithmic time scales in the figures.


\begin{figure}
\begin{center}
\includegraphics[width=8cm, clip]{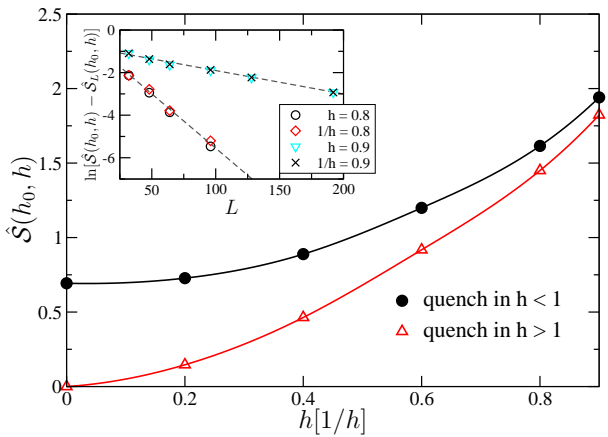}
\caption{(Color online) Saturation values of the entanglement entropy at
$t\to\infty$ and for large $L$.  for quenches from a fully ordered initial
state $h_0\to 0$ to an ordered final sate, as well as from a fully disordered
state $h_0\to\infty$ to another disordered state $h>1$.  The results for
quenches in the region $h<1$ are plotted against $h$, while the results for
quenches in $h>1$ are plotted against $1/h$.   Inset: Finite-size corrections
to the saturation value of the entropy for a quench from $h_0\to 0$ to $h=0.8$ and
$h=0.9$, as well as from $1/h_0\to 0$ to $1/h=0.8$ and $1/h=0.9$.  The asymptotic
values in the large-$L$ limit used for this plot are $\hat{\mathcal
S}(0,0.8)=1.615$, $\hat{\mathcal S}(0,0.9)=1.941$, $\hat{\mathcal
S}(\infty,0.8^{-1})=1.450$ and $\hat{\mathcal S}(\infty,0.9^{-1})=1.823$.
\label{fig3}}
\end{center}
\end{figure}

As seen in the figures, the entanglement entropy for different system sizes $L$
saturates to a value $\hat {\cal S}_L(h_0,h)$ at large $t$. Furthermore, this
saturation value converges to a $L$-independent value for large $L$, i.e.
$\lim_{L \to \infty} \hat {\cal S}_L(h_0,h)=\hat {\cal S}(h_0,h)$. This
asymptotic value for $L\to\infty$ and $t\to\infty$ is larger if the final state
is closer to critical point $h=1$, as shown in Fig.~\ref{fig3}.  Furthermore,
we have obtained an exponential relation for the finite-size correction term,
given by $\hat {\cal S}(h_0,h)-\hat {\cal S}_L(h_0,h) \approx \exp(-L/\xi(h))$,
as illustrated in the inset of Fig.\ref{fig3} for quenches from $h_0\to0$ to
$h=0.8$ and $h=0.9$, as well as for ``dual process'', starting with $1/h_0\to
0$ and ending at $1/h=0.8$ and $1/h=0.9$. From this relation we have estimated
the values of $\xi(h)$ for $h$ close to the critical point, and obtain
$\xi(0.8) \approx \xi(1.25)=19.8(8)$, and $\xi(0.9)\approx \xi(1.11)=95.(5)$,
which are in agreement with the scaling form of the equilibrium correlation
length of the random transverse-field Ising chain:\cite{fisher} $\xi(h) \sim |h-1|^{-2}$.
For this study we have used several initial states
in the ordered and disordered phases, and the same scaling form
for $\xi(h)$ has been found. 


\subsubsection{Quench to the critical point}
\label{sec:global_rand_cr}
\begin{figure}
\begin{center}
\includegraphics[width=8cm, clip]{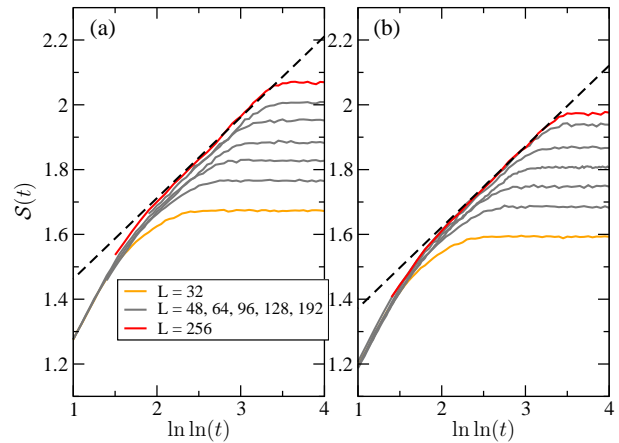}
\caption{(Color online) Average dynamical entropy after a quench to the critical point of the random chain.
(a): The initial state is fully ordered ($h_0\to 0$), (b): the initial state is fully disordered ($1/h_0\to 0$).
The dashed lines in both (a) and (b) have slope 0.25.}
\label{fig4}
\end{center}

\end{figure}

Results for the time dependence of the average entanglement entropy after global quenches
from a fully ordered state ($h_0\to 0$) and a fully disordered state ($1/h_0\to 0$) to 
the critical state are shown in Fig.~\ref{fig4}. 


\begin{figure}
\begin{center}
\includegraphics[width=8cm, clip]{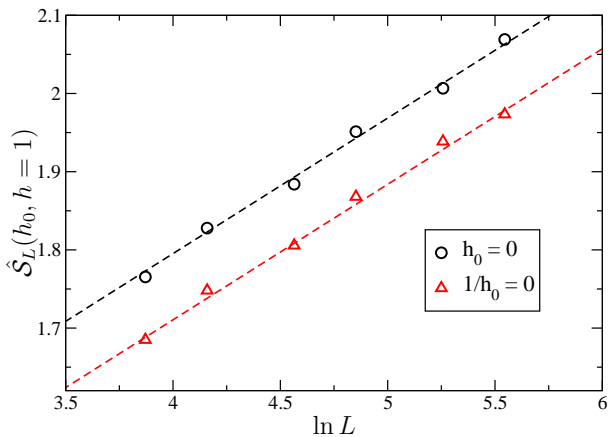}
\caption{(Color online) Saturation value of the average dynamical entropy
as a function of $\ln L$, for quenches from a fully ordered state $h_0\to 0$ and a
fully disordered state $h_0\to\infty$ to the critical point.
The straight dashed lines have the same slope $\ln 2/4$.}
\label{fig5}
\end{center}

\end{figure}

The entanglement entropy for the quench to the critical point increases
with time up to $\tau(L)$, after which it saturates to a value.  The asymptotic
values of the entropy for $t>\tau(L)$ increase monotonously with $L$. Analyzing the
numerical data in Fig.~\ref{fig5} we obtain a logarithmic $L$-dependence: 
\be
\hat {\cal S}_L(h_0,1)=s(h_0)+ b \ln L,
\label{eq:S_asym}
\ee
where the prefactor of the logarithm is
found to be independent of the initial state: $b \approx 0.173 \approx \ln 2/4$;
This is different than the prefactor for the equilibrium entanglement entropy:
$c_{\text{eff}}/3=\ln 2/6$.\cite{refael}

For $t<\tau(L)$, we have found
a double-logarithmic growth of the entropy in time given by 
\be
{\cal S}(t) = s + a \ln\ln t,
\label{eq:dlog}
\ee
with $a\approx 0.25$, which is also independent of the initial state. This
ultraslow growth of the entropy in time reflects the nature of an
infinite-disorder fixed point which is characterized by the activated
dynamics $\ln \tau\sim L^{\psi_{\text{ne}}}$. Combining Eq.~(\ref{eq:S_asym})
and Eq.~(\ref{eq:dlog}), we obtain the exponent $\psi_{\text{ne}}=b/a=0.69(3)$,
which is larger than the exact value, $\psi=1/2$, known for the equilibrium
case. 

We have repeated the calculation for the case when the initial state is also
critical with $h_0=1$. A double-logarithmic growth of $S(t)$ in time and  the
same exponent $\psi_{\text{ne}}$ were obtained.  


\section{Local quench}
\label{sec:local}
In the local quench process, we consider the entanglement entropy of a block
corresponding to one half of the chain with $\ell=L/2$, which is disconnected
from the rest of the chain, with $J_L=J_{L/2}=0$, for $t<0$, and is joined up at
$t=0$ with $J_L=J_{L/2}\neq 0$; the transverse field remains unchanged after the quench.
We study the time evolution of the entanglement entropy for $t>0$. 
\subsection{Homogeneous chain}
\label{sec:local_hom}
\begin{figure}[h!]
\includegraphics[width=8cm, clip]{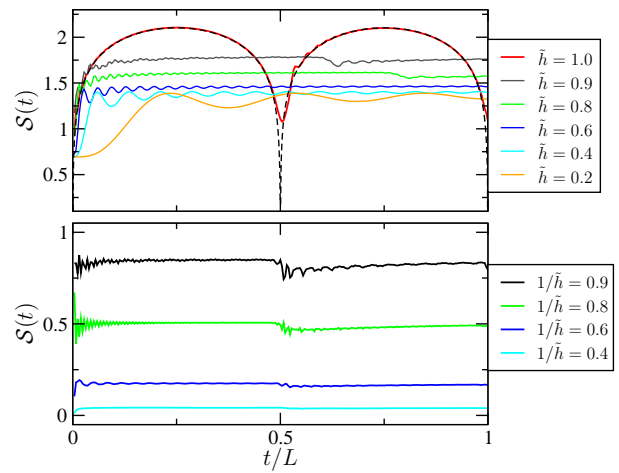}
\caption{\label{fig6}
(Color online) Dynamical entanglement entropy of the homogeneous system
after a local quench versus the rescaled time $t/L$, for $L=256$.
Upper panel: quench to a ordered phase and to the critical point.
For the latter case the conformal result in Eq.~(\ref{eq:loc_form}) is shown by the dashed line.
Lower panel: quench to the disordered phase.}
\end{figure}

For the homogeneous chain of length $L$ the couplings joining the block to the rest of the system for $t>0$
are $J_L=J_{L/2}=1$. We have calculated the entropy dynamics after a
local quench for different values of the transverse field, including the critical value and the values
for ordered as well as disordered phases; the results for $L=256$ are presented
in Fig.~\ref{fig6}.

At the critical point, $\tilde{h}=1$, the entanglement entropy oscillates with a period $t=L/2$,
and this periodic function can be well described as
\be
   {\cal S}(t)=2\frac{c}{3}\ln \left|\frac{L}{2\pi}\sin\frac{2\pi t}{L} \right|+\text{cst.},
   \label{eq:loc_form}
\ee
which was first found in Ref.~\onlinecite{ISzL09}, and has been derived recently
through conformal invariance.\cite{stephan_dubail}

If the quench is performed outside the critical point the dynamical entropy
grows only up to a finite limiting value, as shown in Fig.~\ref{fig6}, both in
the ordered phase (upper panel) and in the disordered phase (lower panel). The
amplitudes of oscillations of $S(t)$ are reduced for large-$L$ and for large
$t$, and the limiting saturation value is of the order of $2\frac{c}{3}\ln
\xi_{\text{hom}}$, with the correlation length, $\xi_{\text{hom}} \simeq
|1-\tilde{h}|^{-1}$, close to the critical point.

\subsection{Random chains}
\label{sec:local_rand}

For a random chain, the couplings are independent random variables 
taken from the uniform distribution in the interval $[0,1]$.
The two couplings $J_L$ and ${J}_{L/2}$ are removed for $t<0$,
and are instantaneously joined to the chain at $t=0$. For the transverse fields 
we use the distribution described in Sec. II. 

\subsubsection{Quench outside the critical point}

\begin{figure}[h!]
\begin{center}
\includegraphics[width=8cm, clip]{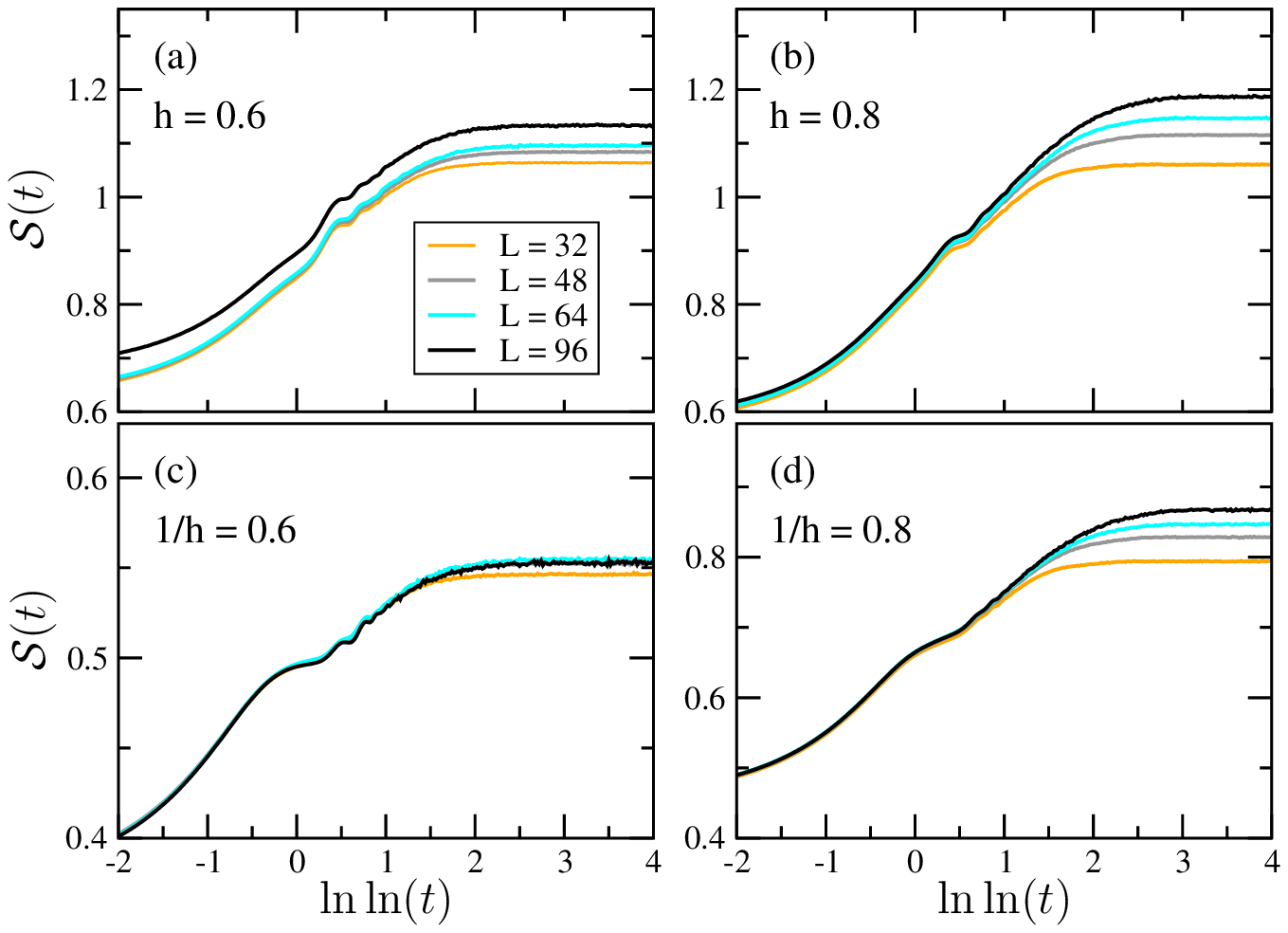}\par
\includegraphics[width=8cm, clip]{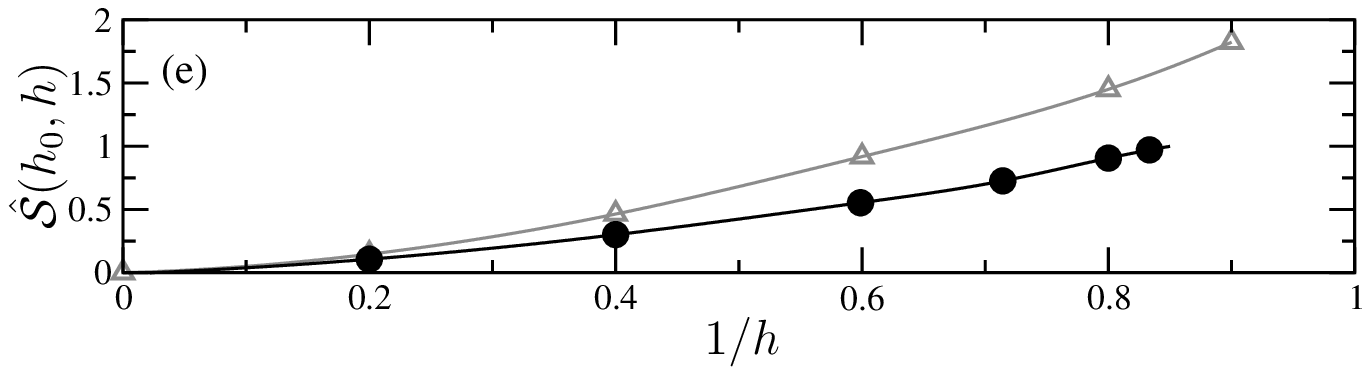}
\caption{(Color online) (a)-(d): Entanglement entropy versus double logarithmic time
after local quenches performed in off-critical phases of the random chain for different lengths. (e): The
data points with filled black circles are the asymptotic
results for large $L$ and large $t$ in the disordered phase. 
The data in gray are results for global quenches in the disordered phase 
(also shown in Fig. \ref{fig3}), given for
comparison.}
\label{fig7}
\end{center}

\end{figure}

We first discuss the time evolution of the disorder-averaged entropy in the non-critical
phases. The results for two examples in the ordered phase are presented in
Fig.~\ref{fig7} (a) and \ref{fig7} (b), and for quenches in the disordered phase are in
Fig.~\ref{fig7} (c) and \ref{fig7} (d).  Due to extremely slow time evolution, the data for
different system sizes $L$ are plotted against $\ln\ln t$.
    

In the long-time regime, the entropy approaches an $L$-dependent saturation value $\hat{\cal S}_L(h)$,
which converges for large sizes: $\lim_{L \to \infty} \hat{\cal S}_L(h)=\hat{\cal S}(h)$.
As shown in Fig.~\ref{fig7}(e), the value $\hat{\cal S}(h)$ increases monotonously as the
critical point is approached; furthermore, this asymptotic value at each $h$ is
smaller than the asymptotic value of the entropy after  a global quench
to an off-critical phase.

\subsubsection{Quench at the critical point}

\begin{figure}[h!]
\begin{center}
\includegraphics[width=8cm, clip]{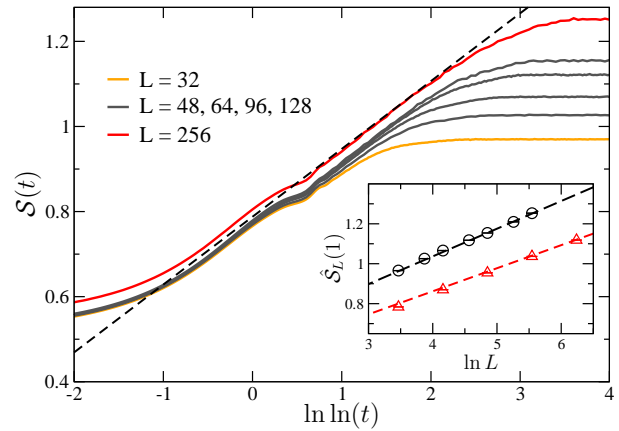}
\caption{(Color online) Disorder-averaged dynamical entropy after a local quench at the critical point of the random chain.
Inset: The asymptotic values at $t\to\infty$ (black circles) for different system sizes, compared with
the static entanglement entropy between two halves of the chain (red triangles). The black dashed
line has slope $\ln 2/5$, and the red dashed line for the static case has slope $\ln 2/6$.   
}
\label{fig8}
\end{center}

\end{figure}

The dynamical entropy after a local quench at the critical point is shown in
Fig.~\ref{fig8} for different lengths of the chain. The overall characteristics
of the entropy in this figure is similar to that in Fig.~\ref{fig4}
obtained after a global quench.

For a fixed length $L$, there is a characteristic time $\tau(L)$, after which
the average entropy is saturated to $\hat{\cal S}_L(1)$. These saturation
values follow a logarithmic $L$-dependence: $\hat{\cal S}_L(1)=s_1+b_1 \ln L$,
for large sizes [see the inset in Fig.~\ref{fig8}]. Here the prefactor of the
logarithm is estimated as $b_1\approx 0.139\approx \ln 2/5$, which is smaller
than the prefactor for a global quench to the criticality: $b\approx \ln 2/4$,
and is slightly larger than the prefactor of the equilibrium entropy:
$c_{\text{eff}}/3=\ln 2/6$.


For $t<\tau(L)$, the average entropy has a double-logarithmic
time-dependence: ${\cal S}(t)=s_1 + a_1 \ln\ln t$ with a prefactor $a\approx0.16(2)$.
Based on the argument for the global quench in sec.~\ref{sec:global_rand_cr},
we obtain $\ln \tau(L) \sim L^{\psi_{\text{ne}}^{\text{loc}}}$ with the
exponent $\psi_{\text{ne}}^{\text{loc}}=0.87(3)$; this exponent is larger
than the exponent $\psi_{\text{ne}}=0.69(3)$, obtained for a global quench.

\section{Discussion}
\label{sec:disc}
We have studied the time-evolution of the entanglement entropy in the random
transverse-field Ising chain after a global and a local quench. The obtained
results are strikingly different from that calculated for the homogeneous version
of the system.  The slow dynamics of entanglement  in a
disordered quantum chain was observed in a previous numerical study,\cite{dyn06}
and has been supported by theoretical work.\cite{burrell_osborne} Our present
numerical study provides clear evidence showing that this dynamics at the
critical point is ultraslow and in a double-logarithmic form.

To explain the difference observed in the time evolution of the entanglement
entropy in homogeneous and in random chains, we apply the semiclassical
quasiparticle picture\cite{CC05} to both cases. In the homogeneous case\cite{semi_class}, the
quasiparticles are related to kinks in the form of domain walls between
differently aligned spin configurations, which are created as entangled fermion
pairs with quasimomenta $\pm p$ moving ballistically in the opposite
directions.  These entangled quasiparticles will contribute to the entanglement
entropy between two regions if one particle arrives in one of the
regions and the other reaches simultaneously the other region. From the
occupation probability of the modes with $p$, one can calculate the
entanglement entropy, as discussed in sec.~\ref{sec:global_hom}, and can
understand the characteristics of the time evolution of the entanglement
entropy.

The properties of the quasiparticles and the dynamics of the entanglement
entropy in the random transverse-field Ising chain can be understood from the
asymptotically exact SDRG. As known from the SDRG, the scaling properties of
the random chain  are described by an infinite-disorder fixed point, at which
disorder fluctuations are completely dominant while quantum fluctuations are
negligible.\cite{fisher,im} The ground state of the random chain consists of a
set of non-overlapping effective spin clusters, each of which has a
characteristic energy scale $\Delta_{\text{cl}}$, given by the excitation
energy of the cluster. The size of a cluster, $\ell_{\text{cl}}$, is
finite in a non-critical phase, its typical value defines the correlation length $\xi$.
(In the ordered phase, there is a giant cluster which is embedded in finite
clusters). At the critical point, where $\xi$ is divergent, for the
largest clusters we have asymptotically: $|\ln \Delta_{\text{cl}}|
\sim \ell_{\text{cl}}^{1/2}$ [cf. Eq.~(\ref{psi})].  The energy-length scaling
described above is also related to Sinai-diffusion in stochastic processes,
\cite{sinai,rand_walk,igloi_rieger98} which explains ultraslow dynamics in
one-dimensional disordered environments.

The spins in a cluster defined in SDRG are maximally entangled. Each cluster
contributes to the entanglement entropy between a block and the rest of the
system by an amount of $s_{\text{cl}}=\ln 2$ as long as it crosses the boundary
of the block.  In equilibrium, the disorder-averaged entanglement entropy at
the critical point, where the correlation length is divergent, is obtained by
summing up contributions of all clusters in the ground state, ${\cal
S}_{\ell}=\sum_{\ell_{\text{cl}}<\ell} s_{\text{cl}}$, yielding ${\cal
S}_{\ell}=\frac{\ln 2}{6}\ln \ell$ for a block of length
$\ell$.\cite{refael} 

In the nonequilibrium case, the cluster formations in the SDRG picture are
time-dependent.  The time span that quantum correlations between different
spins to be built up is the time for the formation of the cluster containing
these spins, and is given by $t_{\text{cl}}\sim \Delta_\text{cl}^{-1}$. The
time span $t_{\text{cl}}$ also corresponds to the time in which a signal emitted at one
end of the cluster arrives at the other end of the cluster via a
Sinai diffusion.  The time-dependent entanglement entropy $S_\ell(t)$ at the
critical point can be obtained by summing over contribution of all entangled
clusters up to time $t$: ${\cal S}_{\ell}(t)=\sum_{t_{\text{cl}}<t}
s_{\text{cl}} \sim \ln \xi(t) \sim \ln \ln t$, where $\xi(t)$ is a
nonequilibrium length-scale, given by $\xi(t) \sim \ln (t)^{1/\psi_{\rm ne}}$.
In the long-time limit when the cluster of size $\ell_{\text{cl}}\lesssim \ell$ are
already formed, the entanglement entropy saturates to a value that is proportional to $\ln \ell$.
The explanation through the SDRG description holds both for global and local
quenches. The main difference between global and local quenches is the excess
energy is finite in a local quench, while it is extensive in a global quench.
When the excess energy is extensive, high energy excitations also contribute to
the dynamics of entangelement entropy and may be responsible for the
nonuniversal short-time behavior; this is similar to the situation in nonrandom
gapless systems after a global quench: conformal field theory describes the
linear growth of the entropy, although excited states strongly influence the
nonequilibrium dynamics.\cite{CC05}
 
The ultraslow behavior of the dynamical entanglement entropy observed in the
random transverse-field Ising chain should be generic for random quantum
systems whose critical point is governed by an infinite-disorder fixed point;
this includes the random $XY$-chain (which has, in fact, an exact mapping
with the Ising chain),\cite{IJ07} the random $XXZ$-chain,\cite{fisher_XX} and
the random quantum Potts chain.\cite{potts} Besides, the quantum criticality of
the random transverse-field Ising model even in higher dimensions is also
controlled by an infinite-disorder fixed point.\cite{2dQMC,2dRG1,2dRG,ddRG,ddRG1,MG12}
It has been found that at this critical point in higher dimensions there is a
singular contribution to the entanglement entropy of the form $\Delta {\cal
S}_{\ell}\sim \log \ell$ for a block in a hypercube form of linear size
$\ell$;\cite{sdrg_entr} this singularity is shown to be related to the
presence of corners. In the dynamical process this corner contribution is
expected to increase in a double-logarithmic fashion. 

\textit{Note added}: After submitting this paper we noticed the preprint
by Levine \textit{et al}\cite{levine12}, in which the time-dependence of the full counting statistics in a disordered
fermion system is studied, which is closely related to the local quench problem
discussed in this work.

\begin{acknowledgments}
We are grateful to I. Peschel and to H. Rieger for discussions.
FI acknowledges support from the Hungarian National Research Fund under grant
No OTKA K75324 and K77629; he
also acknowledges travel support from the NCTS in Taipei, and visitors programs
at National Chengchi University and Academia Sinica.  YCL was supported by the
NSC (Taiwan) under Grant No.~98-2112-M-004-002-MY3.
\end{acknowledgments}

\end{document}